\documentclass[conference]{IEEEtran}
\IEEEoverridecommandlockouts
\usepackage{cite}
\usepackage{amsmath,amssymb,amsfonts}
\usepackage{algorithmic}
\usepackage{graphicx}
\usepackage{textcomp}

\usepackage{xcolor}
\usepackage{subcaption} 

\def\BibTeX{{\rm B\kern-.05em{\sc i\kern-.025em b}\kern-.08em
    T\kern-.1667em\lower.7ex\hbox{E}\kern-.125emX}}

\begin{document}

\title{Community Analysis of Social Virtual Reality Based on Large-Scale Log Data of\\a Commercial Metaverse Platform
}

\author{\IEEEauthorblockN{Hiroto Tsutsui}
\IEEEauthorblockA{
\textit{Institute of Science Tokyo}\\
Tokyo, Japan \\
tsutsui@gs.haselab.net
}
\and
\IEEEauthorblockN{Takefumi Hiraki}
\IEEEauthorblockA{
\textit{University of Tsukuba} \\
Ibaraki, Japan \\
\textit{Cluster Metaverse Lab}\\
Tokyo, Japan \\
hiraki@slis.tsukuba.ac.jp
}
\and
\IEEEauthorblockN{Yuichi Hiroi}
\IEEEauthorblockA{
\textit{Cluster Metaverse Lab}\\
Tokyo, Japan \\
y.hiroi@cluster.mu}
\and
\IEEEauthorblockN{Shoichi Hasegawa}
\IEEEauthorblockA{
\textit{Institute of Science Tokyo}\\
Tokyo, Japan \\
hasevr@haselab.net}
}

\maketitle

\begin{abstract}
This study quantitatively analyzes the structural characteristics of user communities within Social Virtual Reality (Social VR) platforms supporting head-mounted displays (HMDs), based on large-scale log data. By detecting and evaluating community structures from data on substantial interactions (defined as prolonged co-presence in the same virtual space), we found that Social VR platforms tend to host numerous, relatively small communities characterized by strong internal cohesion and limited inter-community connections. This finding contrasts with the large-scale, broadly connected community structures typically observed in conventional Social Networking Services (SNS). Furthermore, we identified a user segment capable of mediating between communities, despite these users not necessarily having numerous direct connections. We term this user segment `community hoppers' and discuss their characteristics. These findings contribute to a deeper understanding of the community structures that emerge within the unique communication environment of Social VR and the roles users play within them.
\end{abstract}

\begin{IEEEkeywords}
Social Virtual Reality, metaverse, community structure, network analysis, community hoppers, bridging users
\end{IEEEkeywords}

\section{Introduction}
\label{intro}
In recent years, Social Virtual Reality (Social VR) platforms supporting head-mounted displays (HMDs) have gained popularity and are hosting new forms of online communities \cite{park2022metaverse}.
Social VR allows users, through avatars, to experience a sense of shared space and physical presence with others (social presence) \cite{Wei2022Communication}.
This study aims to elucidate the patterns of user relationships and the structural characteristics of groups formed within Social VR, thereby deepening our understanding of its communities.

Previous research on online communities has analyzed community structures and user behaviors on various platforms, including SNS, MMORPGs, and others (Sections \ref{sec:related_A}, \ref{sec:related_B}, \ref{sec:related_C}). Regarding Social VR, while the nature of its communities—particularly strong intra-group cohesion (often termed `locality,' indicating dense internal relationships) and the patterns of connections between different groups—has been discussed, these discussions have primarily relied on qualitative methods targeting specific events or small user groups (Section \ref{sec:related_D}). However, comprehensive quantitative analyses of these community structures based on large-scale behavioral log data remain scarce. Consequently, a comprehensive understanding of the general structural characteristics of Social VR communities and their distinctions from other online communities, such as SNS, is still evolving.

To address this gap, this study analyzes community structures based on extensive user behavior log data obtained from Cluster, one of the largest commercial metaverse platforms in Japan. The analysis focuses on substantial user interactions (primarily co-presence time in the same virtual space).
Our analysis quantitatively demonstrates the scale and locality of communities, as well as the connections between them. Furthermore, we identify users who may function as intermediaries between communities and analyze their behavioral characteristics, such as their tendencies to use specific types of virtual spaces. Through these efforts, we aim to provide foundational knowledge on communities in HMD-supporting Social VR. These insights can inform future platform strategies, such as designing features that support both deep engagement within cohesive communities and cross-community interaction, or developing events that mitigate exclusivity, thereby enhancing user experience and community vitality.
\section{Related research}
\label{related}
This section aims to deepen the understanding of community structures in HMD-supporting Social VR by providing a comparative overview of other online platforms like SNS, MMORPGs, and pre-HMD Social VR. While all these platforms feature fluid, user-driven communities, they differ fundamentally in their 
Characteristics of communication, particularly in the balance of synchronous and asynchronous interaction, the primary interaction style (text-based versus voice-based), and the presence or absence of embodied presence. Contrasting these modes with the embodied, synchronous communication central to Social VR helps illuminate how its unique environment might foster distinctive community structures.

\subsection{SNS-based Online Communities}\label{sec:related_A}
In online communities that primarily facilitate interaction through text-based or 2D interfaces, such as SNS and Q\&A sites, a core-periphery structure—comprising a few active core members and many peripheral members—is often observed. While community growth can bring diversity, it can also lead to diluted interactions and reduced cohesion, suggesting that sustainability requires a balance \cite{butler2001membership}. The responsiveness of existing members is crucial for retaining new participants \cite{lampe2005follow}, and in knowledge-sharing sites, a division of labor, where a few contributors primarily support the knowledge base, is also observed \cite{viegas2004newsgroup}.
Network analysis (representing users as nodes and interactions or follow relationships as edges) often reveals that these community connections exhibit small-world properties and scale-free characteristics. This suggests a structure where a few users with exceptionally high degrees connect with many users, thereby exerting significant influence over the entire network \cite{mislove2007measurement}. An important consideration is the potential discrepancy between nominal connections (e.g., friend registrations) and relationship patterns based on actual communication, with the latter often being more limited \cite{wilson2009user}.

\subsection{MMORPGs}\label{sec:related_B}
In MMORPGs like World of Warcraft (WoW), relatively small, persistent groups, commonly known as guilds, are formed. Their size distribution follows a power-law, and they have been reported as fragile social groups characterized by frequent member departures and dissolutions \cite{ducheneaut2007life} (average guild size approximately 16.8 members, median 9). The `Alone Together' phenomenon—where many users are connected simultaneously but predominantly play solo—has also been highlighted, suggesting a disparity between superficial co-presence and relationship patterns based on substantial interactions \cite{ducheneaut2006alone}. Research has shown that community participation increases continued game engagement \cite{seay2004project}, and that role differentiation and communication patterns within guilds correlate with organizational health \cite{ducheneaut2007life}.

\subsection{Pre-HMD Social VR}\label{sec:related_C}
In metaverses like Second Life, where users manipulate 3D avatars via PC screens, diverse user-driven themed communities emerge. However, there is a significant disparity in the popularity of virtual spaces, with users tending to concentrate in certain regions or on specific content \cite{varvello2010exploring}. Friend relationships exhibit small-world properties, but there are also reports that user accounts with an extremely large number of friends are often bots \cite{varvello2010second}. Here too, a discrepancy between nominal relationships and substantial interactions is noted. Users have also been observed to engage with multiple communities, playing a role in information transmission across them.
While these studies utilized large datasets obtained via data crawlers, contemporary Social VR platforms often feature private spaces accessible only to friends, rendering similar data collection methodologies challenging to apply.

\subsection{HMD-supporting Social VR}\label{sec:related_D}
HMD-supporting Social VR, the focus of this study, is characterized by higher immersion, a greater sense of embodiment, and richer non-verbal communication capabilities compared to the pre-HMD Social VR and MMORPGs discussed previously.
As mentioned in the Introduction, the nature of communities in such Social VR, particularly strong intra-group cohesion (`locality') and the patterns of connections between different groups, has, until now, been discussed primarily through qualitative methods such as observations of specific events, interviews with small user groups, and case studies.
For instance, a comprehensive review of Social VR by Maloney et al. \cite{maloney2021social} broadly discusses platform-specific design strategies, communication styles, self-expression, and associated issues like harassment and privacy. This review suggests that Social VR offers unique social experiences and fosters the potential for users to form small communities wherein deep relationships can be built.
Furthermore, research has shown that users can strongly experience a sense of shared space and physical presence with others through avatars (social presence) \cite{sayadi2024feeling}. This experience can form the basis for high-quality interactions comparable to those in the real world (e.g., the potential for rich communication, including non-verbal information, in studies involving older adults \cite{baker2019interrogating}). These studies suggest the potential for building close relationships in Social VR.
Additionally, aspects such as the potential for VR's immersive properties to exacerbate harassment, challenges in platform governance \cite{blackwell2019harassment}, and the cues people utilize in social interactions \cite{fang2021social} have also been investigated.

However, these existing studies often focus on the quality of experience or specific social aspects provided by Social VR, or the depth of interaction within small groups.  

A blog showing community analysis of VRChat \cite{y23586_2018} is still based on snowball sampling and is not comprehensive.
As highlighted in the Introduction, research that comprehensively and quantitatively analyzes the overall community structure of a platform—specifically, the typical size of communities, their internal cohesion, and the strength and patterns of inter-community connections—based on large-scale behavioral log data, remains insufficient.

\section{Analysis}
\label{analysis}
\begin{figure}[htbp] 
  \centering
  \includegraphics[width=0.9\columnwidth]{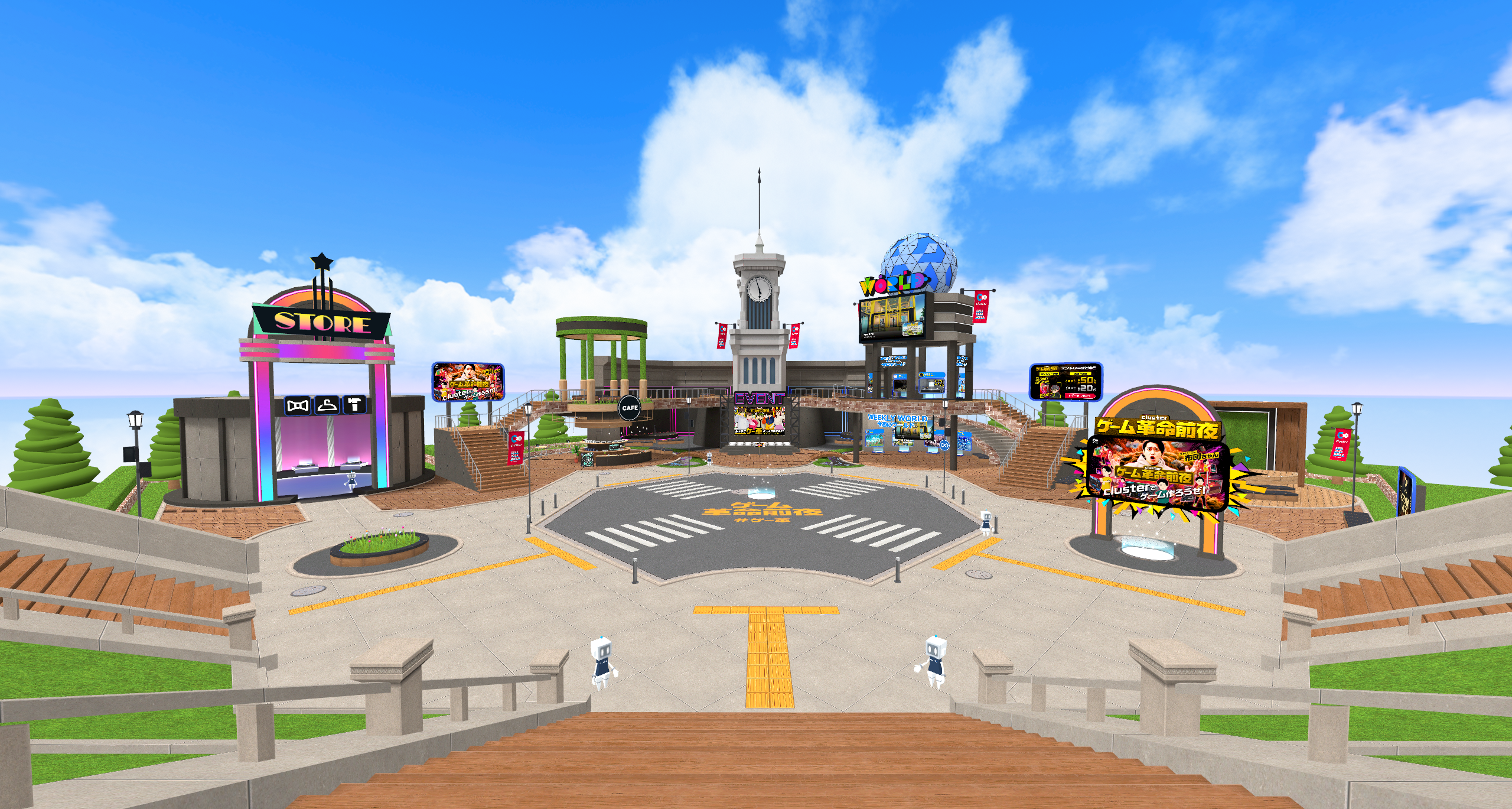}
  \caption{A `Lobby' space in Cluster. All lobbies
share an identical interior design, are constantly accessible,
function as the primary entry point for users, and facilitate
interaction among a diverse range of users.}
  \label{fig:`Lobby' space}
\end{figure}

\begin{figure}[htbp] 
  \centering
  \includegraphics[width=0.9\columnwidth]{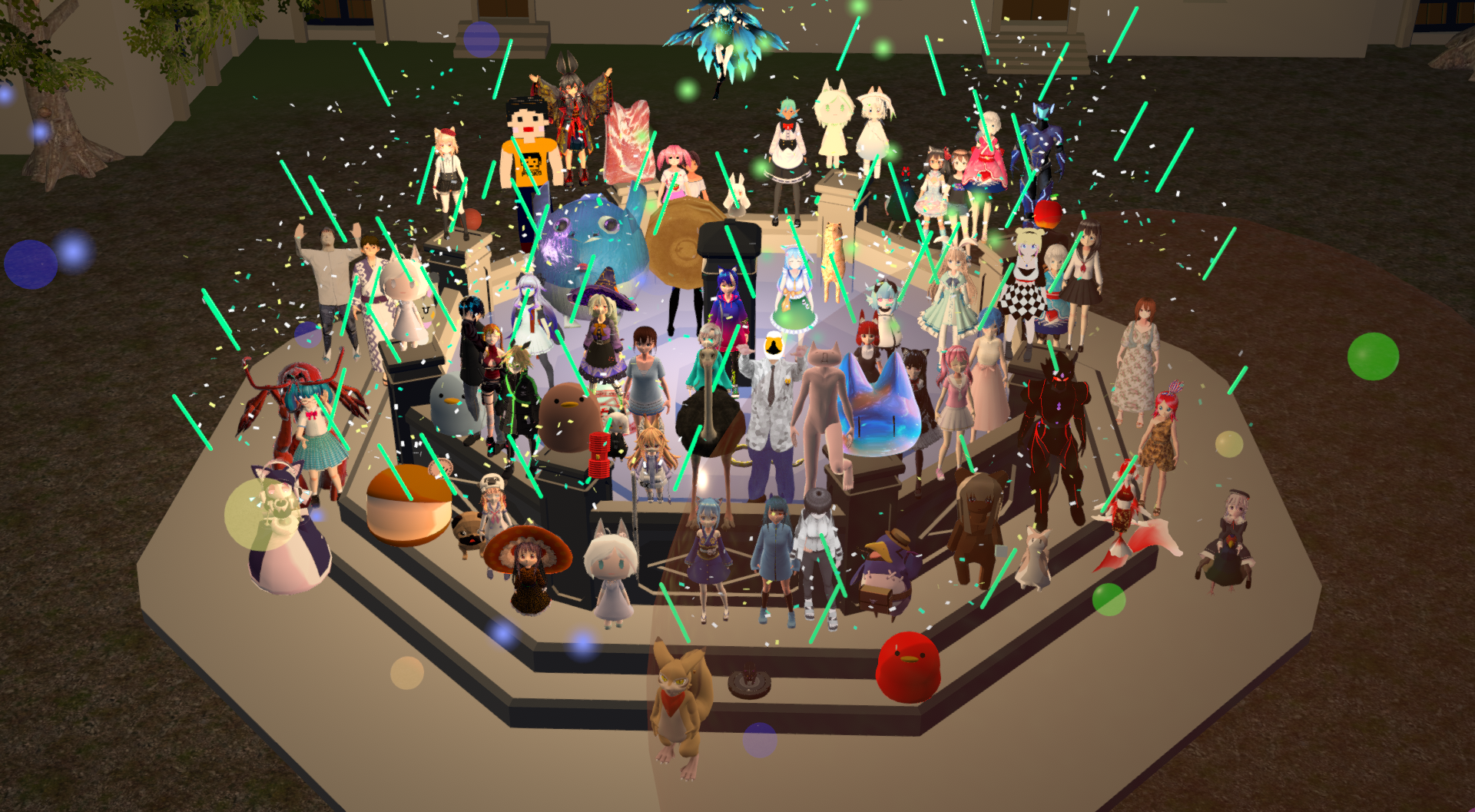}
  \caption{An `Event' space in Cluster. Designed with diverse interiors to suit specific themes, this public space is temporarily available during the event period. It primarily serves as a gathering place for interaction among users interested in that particular theme.}
  \label{fig:`Event' space}
\end{figure}

In this study, we analyze user behavior log data obtained from Cluster, one of the largest commercial metaverse platforms in Japan. Cluster is primarily used by users in Japan and is a multi-platform Social VR platform accessible from various devices, including smartphones, PCs, and head-mounted displays (HMDs). Cluster features virtual spaces, termed `spaces,' which are broadly categorized into `Lobby', `Event', and `World' types. `Lobby' (Fig.\:\ref{fig:`Lobby' space}) is a public space where users are initially guided and where unspecified users can meet. `Event' (Fig.\:\ref{fig:`Event' space}) is a public space for scheduled events with specific themes. `World' refers to all other spaces.
These spaces serve as venues for user interaction, and this study particularly focuses on `Lobby' spaces, which offer broad interaction opportunities, and `Event' spaces.

The analysis proceeds as follows. First, we quantify user relationships based on substantial interactions (defined in this study primarily by concurrent time spent in the same space) and detect/visualize community structures (Section \ref{sec:Understanding community structure}). Next, we quantify the mediating role users play between communities using a `bridging index' and clarify its distribution and characteristics (Section \ref{sec:mediating_user_estimation}). Finally, we analyze the relationship between this bridging index and users' tendencies to use specific space types (Section \ref{sec:room_type_analysis}).
The analysis covers two independent 7-day periods: Period A (March 16, 2025, 15:00 JST -- March 23, 2025, 14:59 JST) and Period B (March 30, 2025, 15:00 JST -- April 6, 2025, 14:59 JST). For each period, the analysis included active users who logged in on 4 or more days per week and were among the top 1,000 in terms of total usage time. The usage times for Period A and Period B were, respectively: median 37 and 34 hours; mean 46 and 43 hours; maximum 164 and 166 hours; and minimum 20 and 18 hours.
The usage data includes user IDs, space entry/exit timestamps (to the minute), space IDs, space type (`World', `Lobby', `Event'), and friend relationships between users.

\subsection{Understanding community structure}\label{sec:Understanding community structure}
\subsubsection{Methodology}\;
To capture substantial interactions between users, we analyzed user relationships based on their `co-presence experience in the same space.' The total co-presence time for a user pair $(i, j)$ was taken as weight $w_{ij}$. The primary analysis focused on thresholded substantial interaction relationships (represented as an undirected graph), where a `connection' was defined if the co-presence time was 2 hours or more. This two-hour threshold was chosen based on preliminary analysis to filter out noise from brief co-presence and extract sustained interactions. It provides the best balance between revealing clear community structures and maintaining a sufficient sample size for analysis.
Following \cite{y23586_2018}, we applied the Infomap algorithm \cite{rosvall2007maps}\cite{rosvall2009map} to the constructed user connection structure to detect communities. Infomap partitions communities based on information compression of random walks. Detected communities were sorted by the number of constituent members. The adjacency matrix representing user connections was then rearranged and visualized, with black indicating a connection present and white indicating its absence.
For comparison, using only Period A data, we also similarly analyzed and visualized: (1) substantial interaction relationships without a co-presence time threshold ($w_{ij} > 0$ defines a connection), and (2) connections based on friend relationships.

\subsubsection{Results}
\begin{figure*}[htbp]
  \centering
  \begin{subfigure}{1.0\columnwidth}
    \centering
    \includegraphics[width=\linewidth]{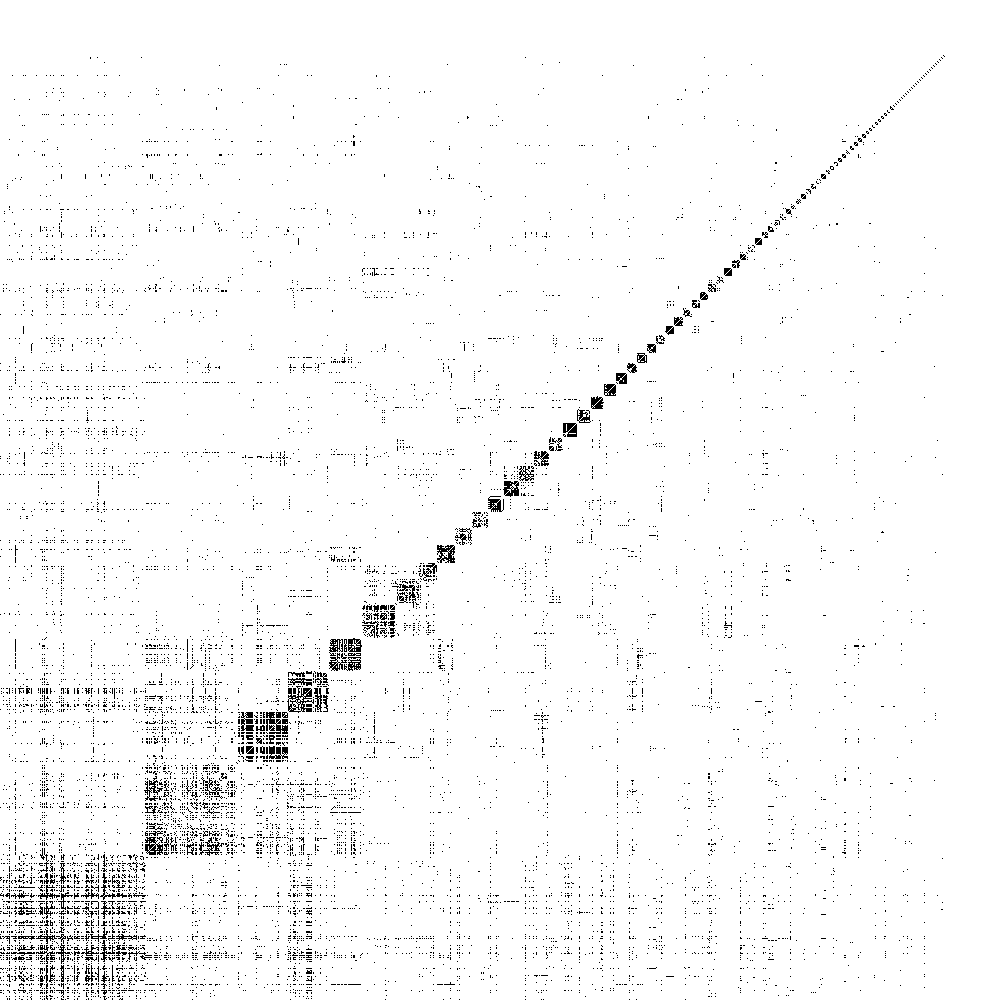}
    \caption{Period A, Substantial interaction (2 hours or more)}
    \label{fig:adj_A_thresh}
  \end{subfigure}
  \hfill
  \begin{subfigure}{1.0\columnwidth}
    \centering
    \includegraphics[width=\linewidth]{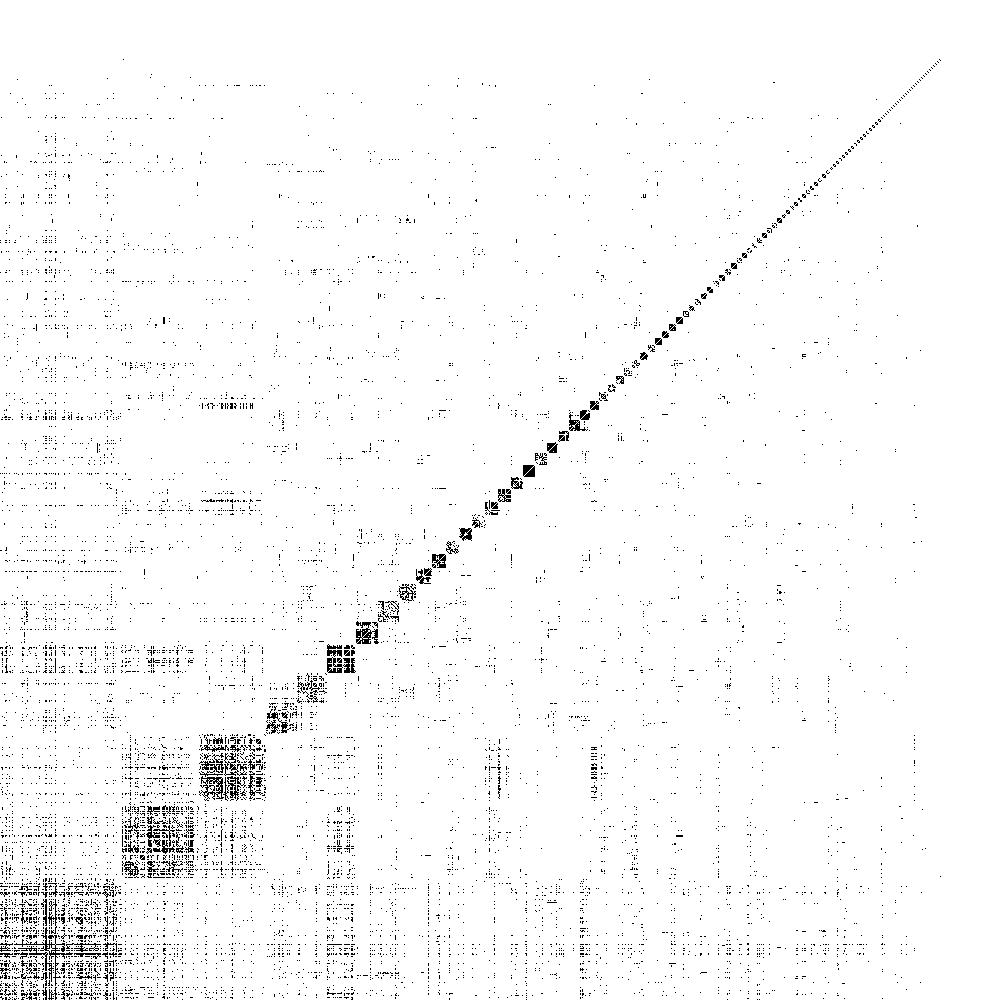}
    \caption{Period B, Substantial interaction (2 hours or more)}
    \label{fig:adj_B_thresh}
  \end{subfigure}
  \vspace{3mm}
  \begin{subfigure}{1.0\columnwidth}
    \centering
    \includegraphics[width=\linewidth]{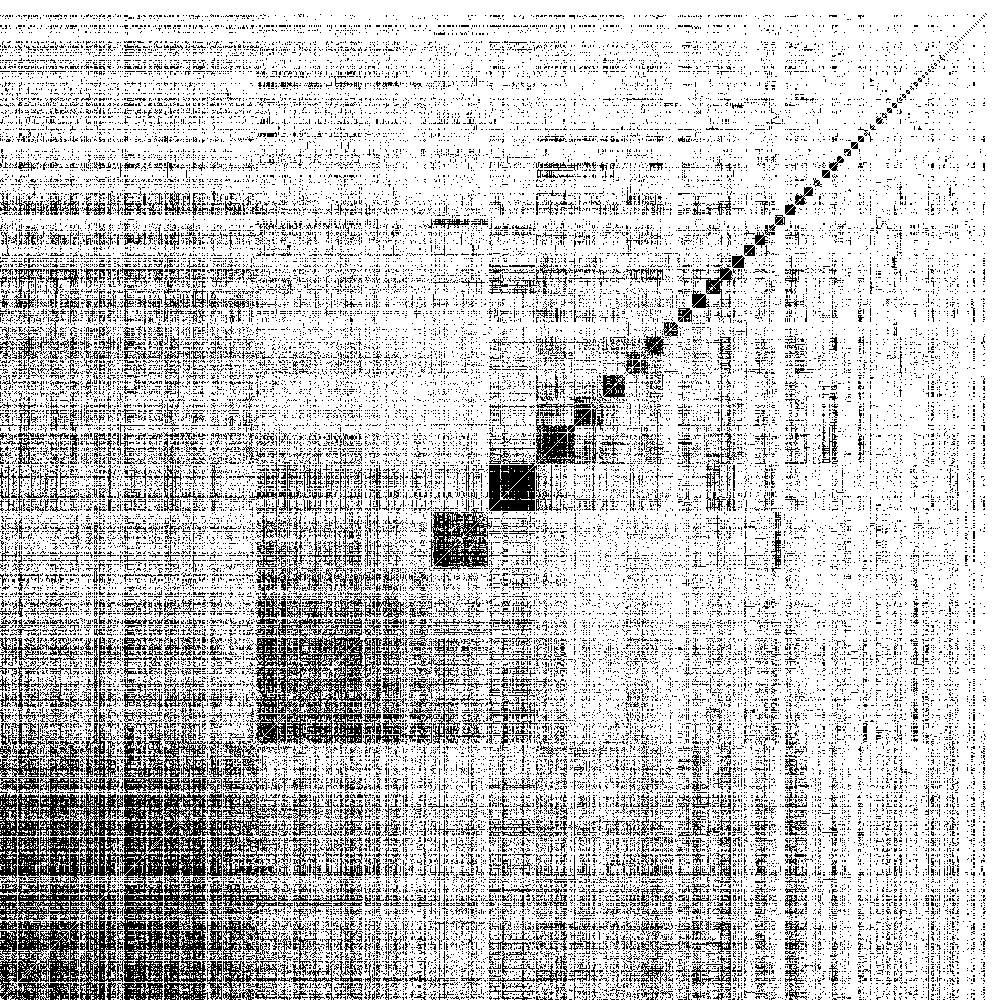}
    \caption{Period A, Substantial interaction (no threshold)}
    \label{fig:adj_A_no_thresh}
  \end{subfigure}
  \hfill
  \begin{subfigure}{1.0\columnwidth}
    \centering
    \includegraphics[width=\linewidth]{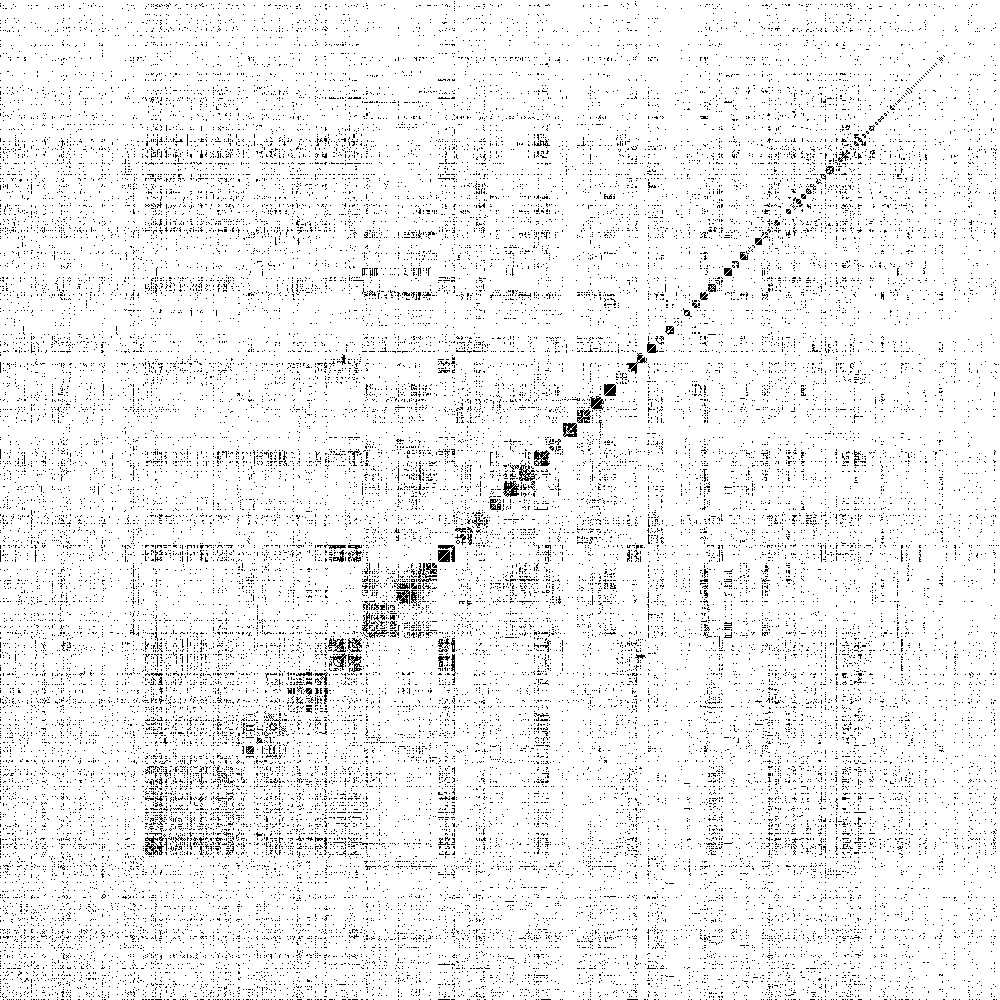}
    \caption{Period A, Friend relationships}
    \label{fig:adj_A_friend}
  \end{subfigure}
  \caption{Adjacency matrices showing user connections under various conditions. Rows and columns in each matrix represent the same users in a common order. Users are sorted by the size of Infomap-detected communities, arranging high-density intra-community connection blocks along the diagonal (bottom-left to top-right). Black pixels indicate user connections; white pixels indicate no connection. (a) and (b) show substantial interactions thresholded at 2 hours or more of co-presence. (c) shows substantial interactions without a co-presence threshold. (d) shows nominal connections based on friend registration, with user order matching (a). Comparing (a) and (c) reveals the effect of thresholding, while comparing (a) and (d) shows the difference between substantial interactions and nominal connections.}
  \label{fig:adjacency_matrices_combined}
\end{figure*}

\;
The visualizations of thresholded substantial interaction relationships (Fig.\:\ref{fig:adjacency_matrices_combined}(\subref{fig:adj_A_thresh}), Fig.\:\ref{fig:adjacency_matrices_combined}(\subref{fig:adj_B_thresh})) showed, for both Period A and B, multiple distinct black blocks along the diagonal (from bottom-left to top-right). These indicate dense connections within detected communities, many of which were relatively small. The portions of the adjacency matrix representing relationships between different communities were predominantly white, indicating that inter-community connections were generally sparse. However, some users did exhibit connections with users from other communities. Overall, these results confirm a structure characterized by many small communities with strong internal cohesion and relatively weak external connections.

When substantial interaction relationships were analyzed without a threshold (Fig.\:\ref{fig:adjacency_matrices_combined}(\subref{fig:adj_A_no_thresh})), larger communities with more ambiguous boundaries were observed compared to the thresholded version (Fig.\:\ref{fig:adjacency_matrices_combined}(\subref{fig:adj_A_thresh})). As this non-thresholded analysis includes relationships based on very short co-presence times, we concluded that the 2-hour threshold was more appropriate for capturing the substantial relationships targeted by this study.

Connections based on friend relationships (Fig.\:\ref{fig:adjacency_matrices_combined}(\subref{fig:adj_A_friend})) were visualized using the same user ordering as that for substantial interaction relationships (Fig.\:\ref{fig:adjacency_matrices_combined}(\subref{fig:adj_A_thresh})) to facilitate comparison. Friend relationships tended to cluster within the segments corresponding to substantial interaction communities. However, in segments representing relatively large communities, the density of friend relationships was sometimes comparatively lower.
In the parts representing relationships between different communities, more connections were observed for friend relationships than for substantial interaction relationships.

\subsection{Estimating mediating users}\label{sec:mediating_user_estimation}
To evaluate user roles within communities, we focus on the `bridging' aspect, which pertains to connections between different communities.

\subsubsection{Methodology}\;
We define a `bridging index' using normalized betweenness centrality \cite{betweeness} to measure the extent to which a user functions as a bridge between different communities. The bridging index (betweenness centrality) $B_v$ for a user $v$ quantifies the degree to which user $v$ lies on the shortest paths connecting other user pairs within the network. Specifically, it is calculated by summing the proportion of shortest paths between all user pairs that pass through user $v$. It is normalized by the theoretical maximum value to yield a value between 0 and 1. In this paper, users with a high bridging index are considered to play a mediating role.

Users' bridging indices were sorted in ascending order and classified into four quartile-based groups (B1: 0--25th percentile, B2: 25th--50th percentile, B3: 50th--75th percentile, B4: 75th--100th percentile). B4 represents the segment with the highest bridging index.
The joint distribution of each user's number of neighbors (number of directly connected users) and bridging index is visualized as a bivariate histogram.

\subsubsection{Results}
\begin{figure*}[htbp]
  \centering
  \begin{subfigure}{1.0\columnwidth}
    \centering
    \includegraphics[width=\linewidth]{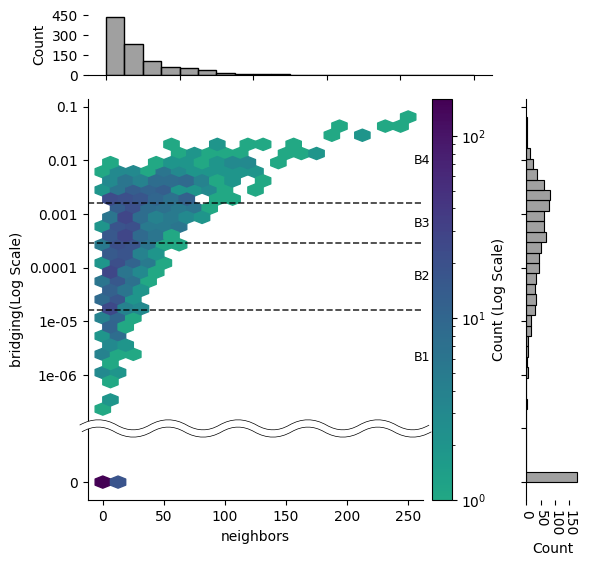}
    \caption{Period A}
    \label{fig:bridging_hist_A_sub}
  \end{subfigure}
  \hfill
  \begin{subfigure}{1.0\columnwidth}
    \centering
    \includegraphics[width=\linewidth]{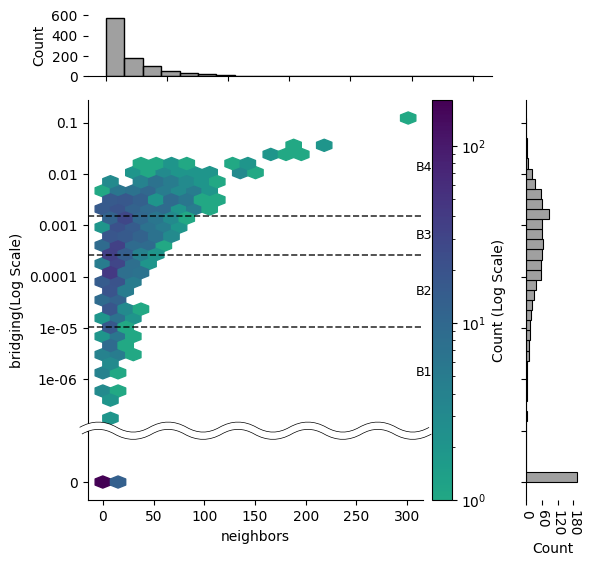}
    \caption{Period B}
    \label{fig:bridging_hist_B_sub}
  \end{subfigure}
  \caption{Bivariate histogram of the number of neighbors (horizontal axis) and bridging index (vertical axis, normalized betweenness centrality, log scale). Color intensity indicates user density. The histogram shows that users with a high bridging index are relatively few. Similar trends are observed in both periods.}
  \label{fig:bridging_hist_combined}
\end{figure*}

\;
The visualization results are shown in Fig.\:\ref{fig:bridging_hist_combined}. In both periods, the bridging index for the vast majority of users was very low, but a segment of users exhibited relatively high values, suggesting they potentially play a mediating role.
Users in the relatively high bridging index segment (approximately 0.02 or higher) generally had a comparatively large number of neighbors. However, for users with a bridging index below this level, the number of neighbors was widely distributed, and a large number of neighbors did not necessarily imply a high bridging index.

\subsection{Analysis of the relationship between space type usage ratios and bridging index}\label{sec:room_type_analysis}
To explore the relationship between user behavioral characteristics and their roles within the community, we calculated each user's usage ratio for each space type (`Lobby', `Event') (time spent in that space type / total time spent in all space types).
We created scatter plots with the number of neighbors on the horizontal axis and `Lobby' usage ratio (or `Event' usage ratio) on the vertical axis, color-coded by bridging index groups (B1 to B4).

\subsubsection{Results}
\begin{figure*}[htbp]
  \centering
  \begin{subfigure}{1.0\columnwidth}
    \centering
    \includegraphics[width=\linewidth]{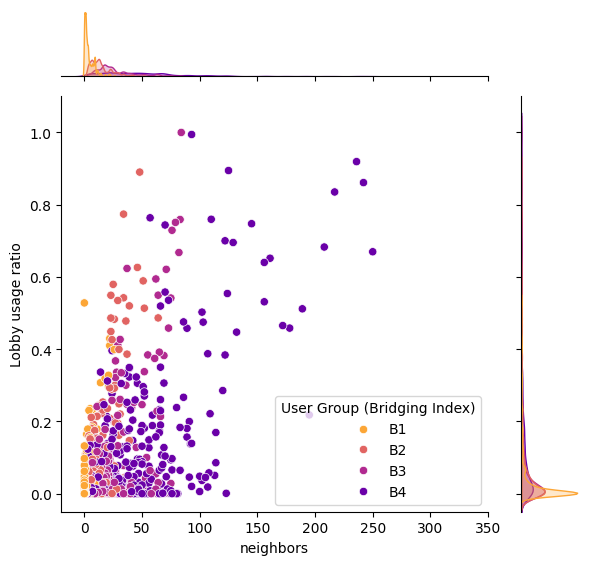}
    \caption{Period A, `Lobby' usage ratio}
    \label{fig:bridging-lobby_scatter_A_sub}
  \end{subfigure}
  \hfill
  \begin{subfigure}{1.0\columnwidth}
    \centering
    \includegraphics[width=\linewidth]{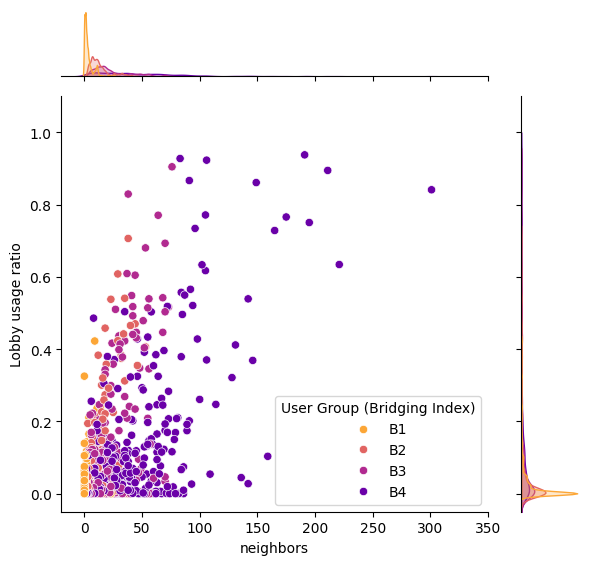}
    \caption{Period B, `Lobby' usage ratio}
    \label{fig:bridging-lobby_scatter_B_sub}
  \end{subfigure}
  \vspace{3mm}
  \begin{subfigure}{1.0\columnwidth}
    \centering
    \includegraphics[width=\linewidth]{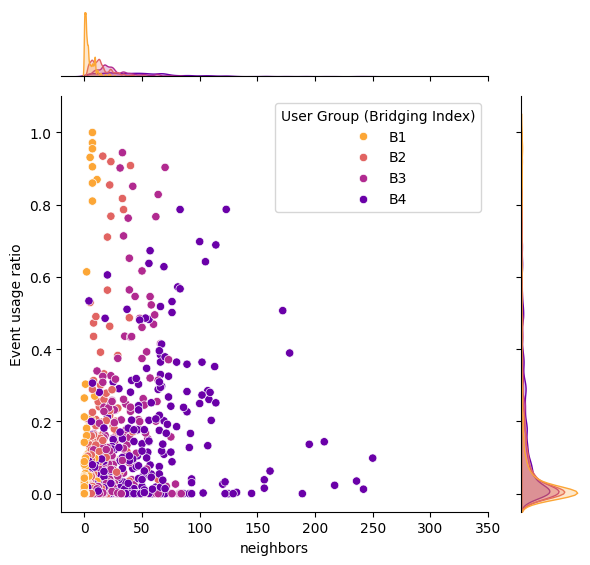}
    \caption{Period A, `Event' usage ratio}
    \label{fig:bridging-event_scatter_A_sub}
  \end{subfigure}
  \hfill
  \begin{subfigure}{1.0\columnwidth}
    \centering
    \includegraphics[width=\linewidth]{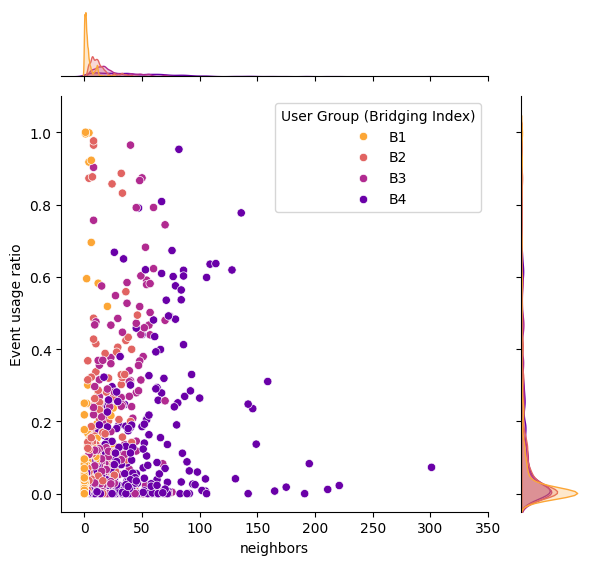}
    \caption{Period B, `Event' usage ratio}
    \label{fig:bridging-event_scatter_B_sub}
  \end{subfigure}
  \caption{Scatter plot of the number of neighbors (horizontal axis) and specific space type usage ratio (vertical axis). Points represent users, color-coded by bridging index quartile groups (B1: lowest to B4: highest). (a),(b) show `Lobby' space usage ratio; (c),(d) show `Event' space usage ratio. B4 users (high bridging index) are more prevalent in segments with high usage ratios for both space types, though the strength of this association varies by space type.}
  \label{fig:room_usage_scatter_combined}
\end{figure*}

\;
The scatter plots (Fig.\:\ref{fig:room_usage_scatter_combined}) illustrate the relationship between `Lobby' and `Event' space usage ratios and the bridging index for each group.
Regarding `Lobby' usage ratio (Fig.\:\ref{fig:room_usage_scatter_combined}(\subref{fig:bridging-lobby_scatter_A_sub}), Fig.\:\ref{fig:room_usage_scatter_combined}(\subref{fig:bridging-lobby_scatter_B_sub})), users in group B4 (highest bridging index) were relatively more prevalent in the high `Lobby' usage ratio segment compared to other groups (B1--B3). However, group B4 also included many users with low `Lobby' usage ratios.
Regarding `Event' usage ratio (Fig.\:\ref{fig:room_usage_scatter_combined}(\subref{fig:bridging-event_scatter_A_sub}), Fig.\:\ref{fig:room_usage_scatter_combined}(\subref{fig:bridging-event_scatter_B_sub})), users in group B4 also tended to be distributed in the high `Event' usage ratio segment compared to other groups, though this trend was less pronounced than for `Lobby' usage ratios. Within group B4, many users had low `Event' usage ratios. Additionally, users in groups with low bridging indices, particularly those with few neighbors, sometimes exhibited relatively high `Event' usage ratios.

\section{Findings}
From the analysis in this study, the following main findings emerged regarding user communities and the characteristics of mediating users on the Social VR platform ``Cluster.''

\subsection{Small-Scale, Cohesive Community Structures and Limited Inter-Community Connections}\;
Analysis of `substantial' interaction relationships (defined by co-presence of 2 hours or more) confirmed that communities in this Social VR environment are typically small-scale with strong internal cohesion, while inter-community connections are limited (Fig.\:\ref{fig:adjacency_matrices_combined}(\subref{fig:adj_A_thresh}), Fig.\:\ref{fig:adjacency_matrices_combined}(\subref{fig:adj_B_thresh})). Furthermore, when no co-presence threshold was applied (Fig.\:\ref{fig:adjacency_matrices_combined}(\subref{fig:adj_A_no_thresh})), more extensive communities with looser connections were observed, indicating that `substantial' relationships in this Social VR environment are formed within a more limited scope. These structures bear similarities to MMORPGs, where guilds form relatively small, persistent groups \cite{ducheneaut2007life}. Conversely, they differ from the large-scale core-periphery structures often observed in SNS—comprising a few active core members and many peripheral members \cite{butler2001membership}—or the presence of central users who connect with numerous other users and wield significant influence \cite{mislove2007measurement}.
Additionally, a comparison with friend relationships (Fig.\:\ref{fig:adjacency_matrices_combined}(\subref{fig:adj_A_friend})) revealed that substantial interactions tend to be more confined to specific groups, whereas friend relationships are formed more broadly. This indicates a difference between nominal connections and substantial interactions, consistent with findings in previous research on other online communities like SNS \cite{wilson2009user}.

\subsection{`Community Hoppers': Existence of a Novel Type of Mediating User}\;
In the Social VR environment studied, prominent `hub' users—who connect with a large number of other users and exert significant influence, as commonly seen in SNS—were not clearly observed. Instead, while some users exhibited high bridging indices, these individuals did not necessarily possess a large number of neighbors (Fig.\:\ref{fig:bridging_hist_combined}). This suggests the existence of users who play a mediating role in `bridging' communities in a manner distinct from merely having a high number of direct connections.
In this paper, we term this user segment—individuals who may play a mediating role between different communities without necessarily possessing an extremely large number of direct connections (i.e., a high number of neighbors)—`community hoppers.' The existence of these `community hoppers' suggests that the forms of influence and centrality in Social VR may differ from those prevalent in conventional online platforms, particularly SNS.

\subsection{Behavioral Characteristics of `Community Hoppers': Usage Trends of Specific Space Types}\;
Users with a high bridging index, potential `community hoppers' (group B4), demonstrated a relatively higher tendency to use highly public space types like `Lobby' and `Event'—where encounters with a diverse range of users are more probable—compared to users with lower bridging indices (especially group B1) (Fig.\:\ref{fig:room_usage_scatter_combined}). However, this was not a simple proportional relationship, as the high bridging index group also included a segment of users with low usage ratios for these space types. Furthermore, users with low bridging indices exhibited a strong tendency towards low `Lobby' space usage (Fig.\:\ref{fig:room_usage_scatter_combined}(\subref{fig:bridging-lobby_scatter_A_sub}), Fig.\:\ref{fig:room_usage_scatter_combined}(\subref{fig:bridging-lobby_scatter_B_sub})), and a similar, albeit less pronounced, trend was observed for `Event' spaces (Fig.\:\ref{fig:room_usage_scatter_combined}(\subref{fig:bridging-event_scatter_A_sub}), Fig.\:\ref{fig:room_usage_scatter_combined}(\subref{fig:bridging-event_scatter_B_sub})). This behavior suggests that some `community hoppers' might function as bridges between different communities by virtue of utilizing spaces where they are likely to encounter a wide variety of users.

\section{Discussion}
The patterns of community connections and user characteristics observed in Social VR in this study may be attributable to the platform's unique communication dynamics. In HMD-supporting Social VR, synchronous interactions via avatars are fundamental, requiring users to share the same time and space. This makes it relatively more difficult to form and maintain shallow, broad connections with a wide range of users at once, compared to SNS where asynchronous communication is predominant. This could contribute to the formation of small, cohesive communities and the less frequent emergence of influential `hub' users connected to many others, unlike what is commonly observed in SNS.
This could lead to the formation of small, cohesive communities with fewer influential 'hub' users than are common on SNS. While this structure fosters a strong sense of belonging, its high cohesion also risks creating barriers for newcomers and promoting insular environments, limiting exposure to diverse information and perspectives.

In such an environment, the existence of `community hoppers' suggests a distinct mechanism for information flow and community linkage in HMD-supporting Social VR. Rather than exerting central influence within a specific community or possessing numerous neighbors, these users may contribute to the diversity of overall community connections and facilitate loose inter-community collaboration by engaging with diverse communities and acting as mediators of information and culture. Granovetter \cite{granovetter1973strength} argued that in social networks, while intimate relationships (strong ties) tend to circulate homogeneous information, less intimate relationships (weak ties) are crucial for bridging different social groups and transmitting novel information and opportunities. The behavior of `community hoppers' identified in this study may contribute to the circulation of information and culture in a similar fashion within the emerging social landscape of Social VR.

Regarding the usage trends of `Lobby' and `Event' spaces, there appears to be an affinity between the `opportunities for encountering diverse others' afforded by these spaces and the behavioral orientations of `community hoppers' (e.g., novelty seeking, information gathering, connecting with others). However, the absence of a simple proportional relationship between usage ratio and bridging index suggests that merely using `Lobby' or `Event' spaces does not inherently lead to a bridging role; rather, users' more active behaviors or intentions are likely key factors. The observation that user groups with low bridging indices also exhibit low `Lobby' space usage ratios might indicate a focus on activities within established communities or a disinclination towards extensive exploratory behavior. For `Event' spaces, it is conceivable that users with low bridging indices might also use them to some extent, especially if the purpose is clear.

The distinction between nominal friend relationships and substantial interactions
indicates that even in Social VR, `acquaintance-level' potential connections are differentiated from `frequently interacting' strong relationships. Substantial interactions are more limited and likely reflect the selectivity of forming dense relationships that involve an investment of time and effort.

These results suggest that in the formation of user relationships within virtual spaces like Social VR, while physical constraints are absent, new factors such as synchronicity, embodiment, and resource allocation (e.g., individual preferences and available time) significantly influence community connection patterns and user roles.

\section{Limitation}
This study has several limitations. First, this study is limited to short-term data from a single platform, Cluster, which is primarily used by Japanese users. Since community structure is heavily influenced by platform design and user culture, the generalizability of these findings to other platforms and diverse user groups requires future comparative validation.
Second, this study relies on user behavior logs (space visit history and friend relationships) and does not directly analyze users' intentions, motivations, or the content and quality of their communication (e.g., conversations). For instance, the actual intentions with which users termed `community hoppers,' utilize spaces and how they interact with others cannot be fully discerned from log data alone.
Third, the choice of analytical methods and parameter settings—such as community detection algorithms, the definition of betweenness centrality, and the co-presence time threshold—can influence the results. While this study employed common methods and seemingly reasonable settings, comparing these results with those from alternative approaches could be a valuable avenue for future work.
Finally, the analysis focused on a relatively active user segment (the top 1,000 users by total usage time), and thus may not fully capture the trends of the entire platform user base, particularly less frequent or newer users.

\section{Conclusion and future work}
This study aimed to quantitatively elucidate the community structure within the `Cluster' Social VR platform—focusing on local community cohesion, inter-community connections, and the characteristics of mediating users—through large-scale log data analysis. The analysis yielded the following main findings: (1) Communities based on substantial interaction (defined as user co-presence for 2 hours or more) in `Cluster' are typically small-scale, exhibit strong internal cohesion, and have limited inter-community connections. (2) Unlike the prominent `hub' users often seen in conventional SNS—who connect with many users and wield significant influence—a distinct group of users, termed `community hoppers,' exists. These users mediate between communities without necessarily possessing numerous direct connections. (3) These `community hoppers' demonstrate a relatively higher tendency to utilize highly public spaces like `Lobby' and `Event'; however, this relationship is not uniform, with the association with `Lobby' space usage being more pronounced.

These findings suggest that the unique communication environment afforded by Social VR (characterized by synchronicity and immersion) fosters user interaction patterns and community dynamics distinct from those of conventional online platforms. The existence of `community hoppers' suggests they may play a novel key role in information dissemination and inter-community collaboration within Social VR.

The metaverse platform ``Cluster'' can be interpreted as offering users the flexibility to choose between `deep interaction within intimate communities' and `broad (or potential) interaction in open spaces.' The interplay between platform design and user behavior is believed to generate the observed community connection patterns and roles such as `community hoppers.'

While this study has shed light on some aspects of community connection patterns and user behavior in Social VR, further research is needed. Future research directions include: (1) verifying the generalizability of the trends identified in this study using data from different Social VR platforms and over longer periods; (2) further investigating the specific behavioral patterns and roles of users, including `community hoppers,' through detailed sequence analysis of behavior logs and comparative analyses between user groups; and (3) conducting time-series analyses of community formation, development, and dissolution. Future work should also complement these quantitative analyses with qualitative research into users' subjective experiences and motivations (e.g., through ethically conducted open-ended surveys or interviews in collaboration with platform operators) to achieve a more multifaceted understanding. Such research is anticipated to contribute to a deeper understanding of Social VR, informing improved platform design and community development strategies.

\section*{Acknowledgment}
This study was supported by JST Moonshot Research \& Development Program Grant Number JPMJMS2013 and JST ASPIRE Grant Number JPMJAP2327, Japan.
We thank the members of the Hasegawa Laboratory and our colleagues for their insightful advice and constructive feedback throughout the research process and on the manuscript. Generative AI (e.g., Google Gemini) was utilized for assistance with English proofreading and refining the language of this paper.

\bibliographystyle{IEEEtran}
\bibliography{IEEEabrv,reference} 

\end{document}